\documentclass[prl,letterpaper,twocolumn,showpacs,superscriptaddress,amsmath,amsfonts,amssymb,floatfix,preprintnumbers]{revtex4}

\usepackage{dcolumn,graphicx,color} 

\definecolor{DarkBlue}{rgb}{0,0,0.5}
\definecolor{White}{rgb}{1,1,1}
\newcommand{\ColorOnline}{\!\!\!\!\!\textcolor{White}{$\blacksquare$}\!(color online).}

\begin{document} \pagestyle{plain}

\title{Relation between the one-particle spectral function and dynamic\\spin susceptibility in superconducting Bi$_2$Sr$_2$CaCu$_2$O$_{8-\delta}$}

\author{D.\,S.~Inosov}\author{S.\,V.~Borisenko}
\affiliation{Institute for Solid State Research, IFW Dresden, P.\,O.\,Box 270116, D-01171 Dresden, Germany.}
\author{I.~Eremin}
\affiliation{Max Planck Institut f\"ur Physik komplexer Systeme, D-01187 Dresden, Germany.} \affiliation{\mbox{Institute
f\"ur Mathematische und Theoretische Physik, TU-Braunschweig, 38106 Braunschweig, Germany.}}
\author{A.\,A.~Kordyuk}
\affiliation{Institute for Solid State Research, IFW Dresden, P.\,O.\,Box 270116, D-01171 Dresden, Germany.}
\affiliation{Institute of Metal Physics of National Academy of Sciences of Ukraine, 03142 Kyiv, Ukraine.}
\author{V.\,B.~Zabolotnyy}\author{J.~Geck}\author{A.~Koitzsch}\author{J.~Fink}\author{M.~Knupfer}\author{B.~B\"uchner}
\affiliation{Institute for Solid State Research, IFW Dresden, P.\,O.\,Box 270116, D-01171 Dresden, Germany.}
\author{H.~Berger}
\affiliation{Institut de Physique Appliqu\'{e}e, Ecole Politechnique F\'{e}derale de Lausanne, CH-1015 Lausanne,
Switzerland.}
\author{R.~Follath}
\affiliation{BESSY GmbH, Albert-Einstein-Strasse 15, 12489 Berlin, Germany.}

\begin{abstract}

Angle resolved photoemission spectroscopy (ARPES) provides a detailed view of the renormalized band structure and,
consequently, is a key to the self-energy and the single-particle Green's function. Here we summarize the ARPES data
accumulated over the whole Brillouin zone for the optimally doped Bi$_2$Sr$_2$CaCu$_2$O$_{8-\delta}$ into a parametric
model of the Green's function, which we use for calculating the itinerant component of the dynamic spin susceptibility
in absolute units with many-body effects taken into account. By comparison with inelastic neutron scattering (INS) data
we show that the itinerant component of the spin response can account for the integral intensity of the experimental INS
spectrum. Taking into account the bi-layer splitting, we explain the magnetic resonances in the acoustic (odd) and optic
(even) INS channels.

\end{abstract}




\maketitle

\noindent The origin of the magnetic resonance structure observed in the superconducting (SC) state of
YBa$_2$Cu$_3$O$_{6+\delta}$ (YBCO) \cite{CollectionYBCO, HaydenNature04, PailhesSidis04, PailhesUlrich06, WooNature06},
Bi$_2$Sr$_2$CaCu$_2$O$_{8-\delta}$ (BSCCO) \cite{FongBourges99, HeSidisBourges00, CapognaFauque06}, and other families
of cuprates \cite{TranquadaWoo04} is one of the most controversial topics in today's high-$T_c$ superconductor (HTSC)
physics. Existing theories waver between the itinerant magnetism resulting from the fermiology \cite{CollectionRPA,
LiuZhaLevin95, AbanovChubukov99, EreminMorr05, EreminMorr06} and the local spins pictures (such as static and
fluctuating ``stripes'', coupled spin ladders, or spiral spin phase models) \cite{TranquadaWoo04, Tranquada05,
CollectionLoc}, as it appears that both approaches can qualitatively reproduce the main features of the magnetic spectra
in the neighborhood of the optimal doping. It is a long standing question, which one of these two components (itinerant
or localized) predominantly forms the integral intensity and the momentum-dependence of the magnetic resonances. It is
therefore essential to estimate their contribution quantitatively, carefully taking into account all the information
about the electronic structure available from experiment. However, such a comparison, which could shed light on the
dilemma, is complicated, as it requires high-quality INS data and the extensive knowledge of the electronic structure
for the same family of cuprates. On the other hand, APRES data for YBCO compounds, for which the best INS spectra are
available, are complicated by the surface effects \cite{Zabolotnyy06}, while for BSCCO, most easily measured by
surface-sensitive techniques such as ARPES, the INS measurements show much lower resolution due to small crystal sizes.

Here we propose a way to estimate the dynamic spin susceptibility in the odd (o) and even (e) channels within the random
phase approximation (RPA) from the single-particle spectral function, including many-body effects, and compare the
resulting spectrum calculated for optimally doped BSCCO with the available INS measurements on both BSCCO and YBCO.

We start with establishing the relation between the quasiparticle Green's function and INS response. The normal-state
Lindhard function is related to the quasiparticle Green's function via the following summation over Matsubara
frequencies \cite{BruusFlensberg, MonthouxPines94}:\vspace{0.1em}
\begin{equation}
\chi_0(\textbf{Q},i\Omega_n)\kern-1pt=\smash{\frac{1}{\pi^2}\kern-2pt\int\sum_m}\,
G(\textbf{k},i\omega_m)\,G(\textbf{k}+\textbf{Q},i\omega_m+i\Omega_n)\,d\textbf{k}\vphantom{\sum}
\label{EqMatsubara}\vspace{0.25em}
\end{equation}

\noindent Besides the bare Green's function, equation (\ref{EqMatsubara}) also holds for the renormalized one. It can be
rewritten as a double integral along the real energy axis \cite{DahmTewordt95,EschrigNorman03}:\vspace{0.1em}
\begin{equation}
\chi_0^{\textup{o},\textup{e}}(\textbf{Q},\Omega)=\smash{\kern-4pt\sum_{\lower1em\hbox{$\stackrel{\scriptstyle
i=j\,\textup{(o)}}{\scriptstyle i\neq j\,\textup{(e)}}$}}\kern-2pt\iint\limits_{\,-\infty}^{\,~~~~+\infty}\kern-3pt}
C_{ij}(\textbf{k},\epsilon,\nu)~
\frac{n_f(\nu)-n_f(\epsilon)}{\Omega+\nu-\epsilon+i\,\Gamma}~d\nu\,d\epsilon,\label{EqHi0}\vspace{0.6em}
\end{equation}
\noindent where
$C_{ij}(\textbf{k},\epsilon,\nu)=\frac{1}{\pi^{2^{\vphantom{0}}}}\textstyle{\int}\operatorname{Im}G_i(\textbf{k},\epsilon)\operatorname{Im}G_j(\textbf{k}+\textbf{Q},\nu)\,d\textbf{k}$
is the cross-correlation of the constant-energy cuts of the spectral function over the Brillouin zone (BZ),
$n_f(\epsilon)=1/(e^{\epsilon/k_BT}+1)$ is the Fermi function, and indices $i$ and $j$ numerate the bonding and
antibonding bands. The factors $C_{ij}(\textbf{k},\epsilon,\nu)$ can be efficiently calculated in the Fourier domain by
means of the cross-correlation theorem \cite{Papoulis62}.

\begin{figure*}[t]
\includegraphics[width=\textwidth]{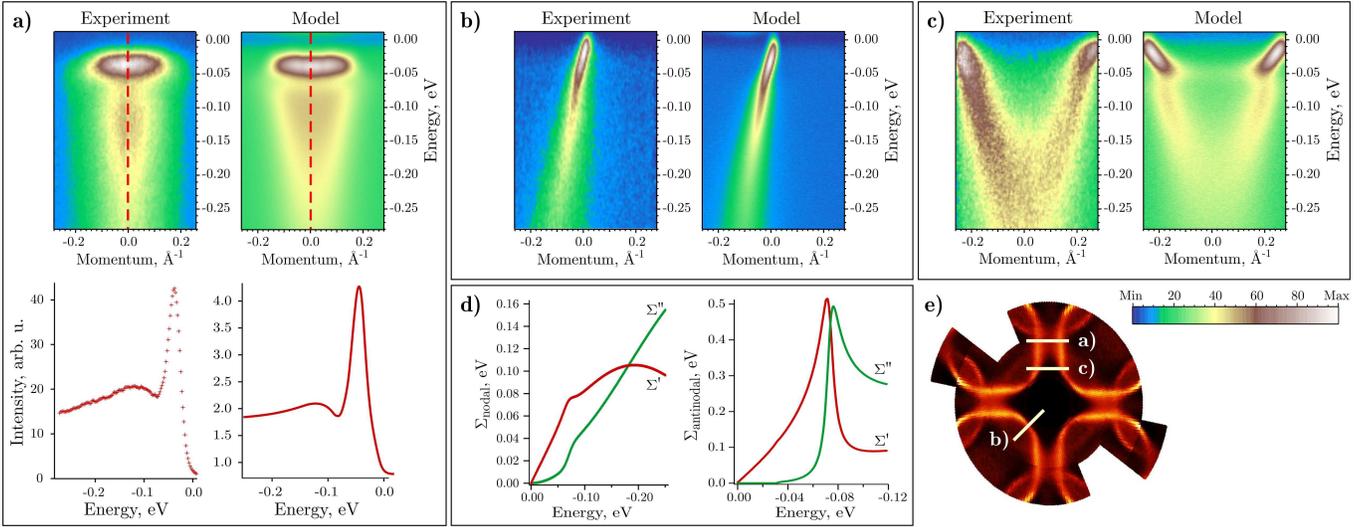}
\caption{\kern0.2pt\ColorOnline~Comparison of the model with experimental ARPES spectra of OD Bi-2212 at 30~K taken with
50~eV (a,~b) and 38~eV (c) photon energies. The model spectra are smoothed with a gaussian to account for 20\,meV energy
resolution and 0.025\,\AA$^{-1}$ angular resolution. \textbf{a)}~Spectra at the $(\pi,0)$ point with the corresponding
energy distribution curves (below) taken along the dashed lines. \textbf{b)}~Nodal spectra along the $(\pi,\pi)$
direction. \textbf{c)}~Comparison of the experimental and model spectra taken at an intermediate position in $k$-space
to check the validity of the interpolation of the self-energy between the nodal and antinodal directions.
\textbf{d)}~The Kramers-Kronig consistent real and imaginary parts of the nodal and antinodal self-energies. \textbf{e)}
Positions of the cuts a)\,--\,c) in $k$-space.}\vspace{-0.7em} \label{f:Model}
\end{figure*}

In the SC state, the anomalous Green's function $F$ additionally contributes to $\chi_0$
\cite{AbanovChubukov99}:\vspace{-0.7em}
\begin{multline}
C_{ij}(\textbf{k},\epsilon,\nu)=\frac{1}{\pi^2}
\kern-2pt\int\bigl[\operatorname{Im}G_i(\textbf{k},\epsilon)\operatorname{Im}G_j(\textbf{k}+\textbf{Q},\nu)\\
+\operatorname{Im}F_i(\textbf{k},\epsilon)\operatorname{Im}F_j(\textbf{k}+\textbf{Q},\nu)\bigr]\,d\textbf{k}
\vspace{-0.6em}
\end{multline}
Although $\operatorname{Im}\kern-1pt F$ is not directly measured by ARPES, it can be still accounted for, as we will
subsequently show.

After one knows the Lindhard function $\chi_0$ (frequently referred to as the bare spin susceptibility), one can finally
obtain from RPA the dynamic spin susceptibility $\chi$ \cite{LiuZhaLevin95}, the imaginary part of which is directly
proportional to the measured INS intensity \cite{Tranquada05}:\vspace{-0.2em}
\begin{equation}
\chi^{\textup{o},\textup{e}}(\textbf{Q},\Omega) =
{\chi_0^{\textup{o},\textup{e}}(\textbf{Q},\Omega)}/{\bigl[1-J_Q^{\,\textup{o},\textup{e}}\,\chi_0^{\textup{o},\textup{e}}(\textbf{Q},\Omega)\bigr]}
\label{EqRPA}\vspace{-0.2em}
\end{equation}

The coefficient $J_Q^{\,\textup{o},\textup{e}}$ in the denominator of (\ref{EqRPA}) describes the effective Hubbard
interaction. In our calculations we employed the model for $J_Q^{\,\textup{o},\textup{e}}$ discussed in
\cite{EreminMorr06,BrinckmannLee01}, namely:
\begin{equation}
J_Q^{\,\textup{o},\textup{e}}=-J_\|(\cos{Q_x}+\cos{Q_y})\pm J_\perp,
\end{equation}
where the first term accounts for the $Q$-dependence due to the in-plane nearest-neighbor superexchange, and the second
term arises from the out-of-plane exchange interaction.

Thus, knowing the single-particle Green's function leads us to a comparison of ARPES results with the INS data. The
previous calculations based on this idea \cite{AbanovChubukov99, EreminMorr05, EreminMorr06} were performed for the bare
band structure only, disregarding the renormalization effects, which makes the conclusions based on comparison with the
INS data rather uncertain. The recent work by U.\,Chatterjee \textit{et~al.} \cite{Chatterjee06} is the only available
paper that includes the many-body effects from experimental data (in a procedure different from ours), but it does not
account for the bi-layer splitting (necessary for reproducing the odd and even INS channels), provides the results in
arbitrary units only, rather than on an absolute scale, and gives only an estimate for the anomalous contribution to
$\chi$. So we will address these issues in more detail below.

At first, we will introduce an analytical model that can reproduce the ARPES measurements within a wide energy range and
all over the BZ. As in a single experiment it is practically impossible to obtain a complete data set of ARPES spectra,
such a model allows making use of all the available data measured from a particular sample and calculating the full 3D
data set afterwards. In such a way the effect of matrix elements and experimental resolution is also excluded.

The measured ARPES intensity is basically proportional to the imaginary part of the Green's function (although it is
affected by matrix elements, experimental resolution, and other factors \cite{Borisenko01}). The latter can be obtained
if one knows the self-energy, extracted from the ARPES data in a routine self-consistent Kramers-Kronig procedure
\cite{Kordyuk}.

We employed a model of the Green's function based on the bare electron dispersion studied in \cite{Kordyuk} and a model
for the imaginary part of the self-energy $\Sigma'' = \Sigma''_\textup{el} + \Sigma''_\textup{bos}$, where
$\Sigma''_\textup{el} = \alpha\kern1pt\omega^2$ is the Fermi-liquid component of the scattering rate that originates
from the electron-electron interactions, and $\Sigma''_\textup{bos}$ models the coupling to a bosonic mode
\cite{RemSelfEnergy}. In the $(\pi,\pi)$ (nodal) direction we modeled $\Sigma''_\textup{bos}$ by a step-like function
$\Sigma''_\textup{bos}\kern-4pt=\kern-2pt
\frac{1}{2}\,\beta_\textup{n}\bigl[1\kern-2pt+\kern-1pt\tanh\bigl(\frac{-\omega-\Omega_\textup{n}}
{\delta\omega_n}\bigr)\bigr]$ of width $\delta\omega_\textup{n}$, height $\beta_\textup{n}$ and energy
$\Omega_{\textup{n}}$, while in the $(\pi,0)$ (antinodal) direction we accounted for the peak in the self-energy due to
the pile-up in the density of states at the gap energy: $\Sigma''_\textup{bos}\kern-3pt=\kern-1pt
-\beta_\textup{a}\,\textup{Re}\, \frac{\omega}{\kern-2pt\sqrt{(\omega\kern1pt-\kern1pt
i\,\delta\omega_\textup{a})^2\kern1pt-\kern1pt(\Delta_0\kern1pt+\kern1pt\Omega_\textup{a})^2}}$, where $\Delta_0$ is the
SC gap, $\Omega_\textup{a}$ is the mode energy, and $\delta\omega_\textup{a}$ is the broadening parameter (see
\cite{SacksCren06} and references therein). The real part of the self-energy $\Sigma'$ was then derived by the
Kramers-Kronig procedure and the Green's function was calculated according to \cite{ChubukovNorman04}:\vspace{-0.5em}
\begin{equation}
G(\textbf{k},\omega)\kern-1pt=\kern-1pt\frac{\omega-\Sigma(\textbf{k},\omega)+\epsilon_\textbf{k}}
{\bigl[\omega\kern-1pt-\kern-1pt\Sigma(\textbf{k},\omega)\bigr]^2\kern-2.5pt-\Delta^2(\textbf{k})\Bigl[1\kern-1pt-\frac{\Sigma(\textbf{k},\omega)}{\omega}\Bigr]^2
\kern-3pt-\epsilon_\textbf{k}^2},\vspace{-0.5em} \label{EqChubukov}
\end{equation}
\noindent where $\Delta(\textbf{k})$ is the SC d-wave gap changing from zero along the BZ diagonals to the maximal value
of $\pm\Delta_0$ along the antinodal directions. Self-energy parameters were specified independently for the nodal and
anti-nodal parts of the spectra, with a d-wave interpolation between these two directions: $\Sigma''(\textbf{k},\omega)
= \Sigma''_\textup{n}(\omega) + \frac{1}{4}\bigl[\Sigma''_\textup{a}(\omega) - \Sigma''_\textup{n}(\omega)\bigr](\cos
k_x-\cos k_y)^2$. We also assumed the particle-hole symmetry in $\Sigma''$. To achieve the best reproduction of the
experimental data, all the free parameters were adjusted during comparison with a set of ARPES spectra of Bi-2212 to
achieve the best correspondence (Fig.\,\ref{f:Model}). The best-fit parameters of the model are listed in the following
table:\vspace{-0.2em}
\begin{center}
\noindent \footnotesize \begin{tabular}{llll}\hline\vspace{-0.75em}\\ \smallskip
$\alpha = 3.0$\,eV$^{-1}$ & $\beta_\textup{n} = 30$\,meV & $\beta_\textup{a} = 200$\,meV & $\delta\omega_\textup{n} = 10$\,meV \\
\smallskip
$\delta\omega_\textup{a} = 0.08\,\Delta_0$ & $\Omega_\textup{n} = 60$\,meV & $\Omega_\textup{a} = 42$\,meV & $\Delta_0 = 35$\,meV  \\
\hline
\end{tabular}
\end{center}\normalsize\vspace{-0.2em}

We would like to stress here that such a simple self-energy model that includes coupling only to a single bosonic mode
can accurately reproduce the state of the art ARPES spectra of BSCCO, as we have just shown.

\begin{figure}[t]
\includegraphics[width=\columnwidth]{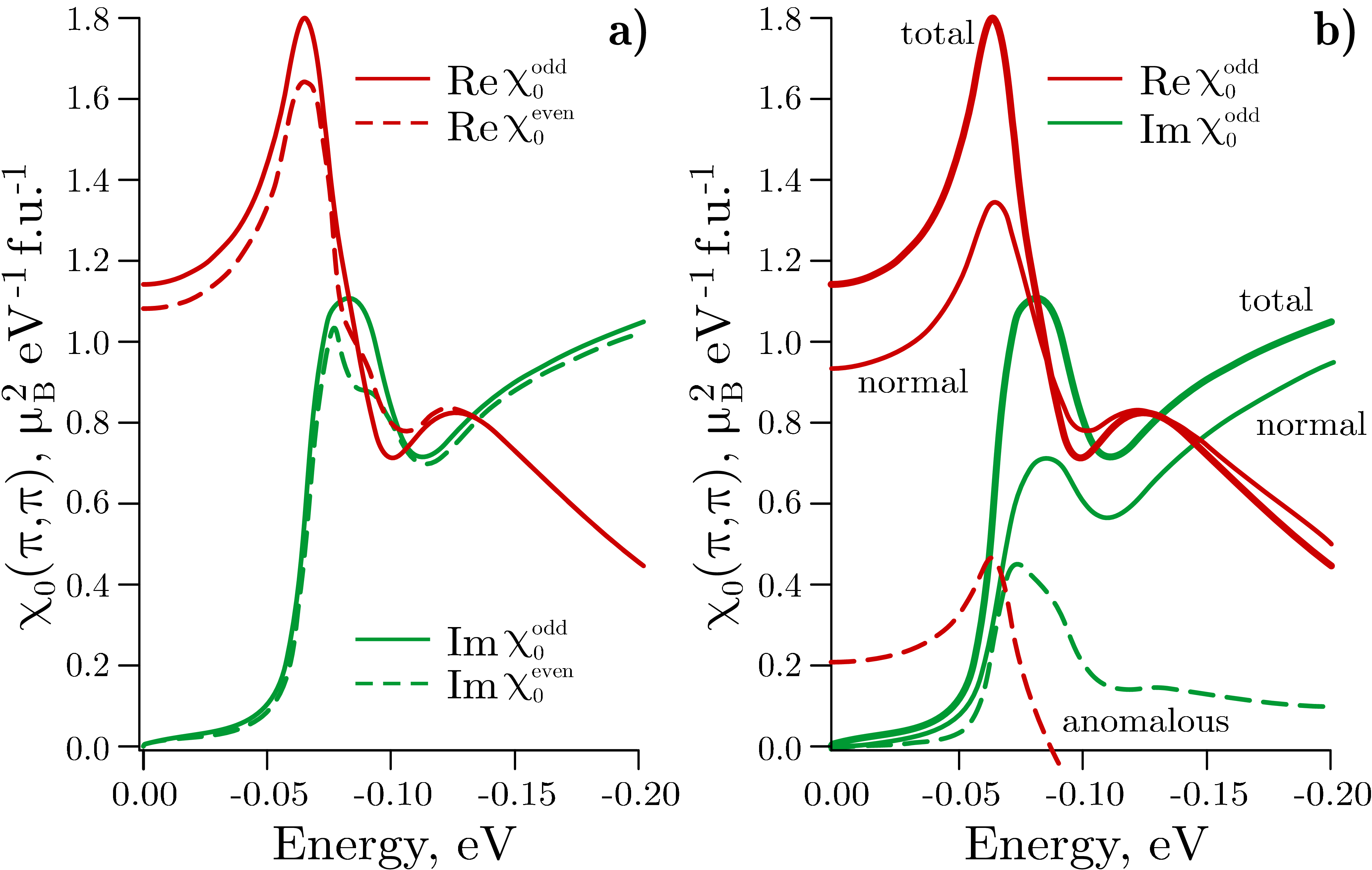} \caption{\ColorOnline~\textbf{a)}~Energy dependence of the real and
imaginary parts of the Lindhard function $\chi_0$ at the $(\pi,\pi)$ point for the odd and even channels.
\textbf{b)}~Contributions of the normal (thin solid curves) and anomalous (dashed curves) components to the real and
imaginary parts of $\chi_0^\textup{odd}$ in the SC state. The sum of two components is shown as thicker curves. In our
calculations we used the $\Gamma$ value in (\ref{f:Chi0}) of 5~meV, which could introduce insignificant additional
broadening of $\chi_0$ as compared to the bare band calculations. The energy integration range in (\ref{EqHi0}) was
chosen to be $\pm$0.25~eV.} \label{f:Chi0}
\end{figure}

The described model has multiple advantages for numerical calculations over raw ARPES data. Only in such a way one can
completely separate the bonding and antibonding bands, which is impossible to achieve in the experiment, and reveal the
nature of the odd and even channels of the two-particle spectrum. With the self-energy $\Sigma(\textbf{k},\omega)$ and
the pairing vertex $\Delta(\textbf{k})\Bigl[1-\frac{\Sigma(\textbf{k},\omega)}{\omega}\Bigr]$ present in the model, it
becomes possible to calculate the anomalous Green's function $F(\textbf{k},\omega)$
\cite{HaslingerChubukov03}:\vspace{-0.5em}
\begin{equation}
F(\textbf{k},\omega)\kern-1pt=\kern-1pt\frac{\Delta(\textbf{k})\Bigl[1\kern-1pt-\frac{\Sigma(\textbf{k},\omega)}{\omega}\Bigr]}
{\bigl[\omega\kern-1pt-\kern-1pt\Sigma(\textbf{k},\omega)\bigr]^2\kern-2.5pt-\Delta^2(\textbf{k})\Bigl[1\kern-1pt-\frac{\Sigma(\textbf{k},\omega)}{\omega}\Bigr]^2
\kern-3pt-\epsilon_\textbf{k}^2}\vspace{-0.5em} \label{EqAnomalous}
\end{equation}
Besides the already mentioned absence of matrix element effects and experimental resolution, the formulae
(\ref{EqChubukov}) and (\ref{EqAnomalous}) also allow to obtain both real and imaginary parts of the Green's functions
for all $\textbf{k}$ and $\omega$ values including those above the Fermi level. It automatically implies the
particle-hole symmetry ($\epsilon_{\textbf{k}_f-\textbf{k}} = -\epsilon_{\textbf{k}_f+\textbf{k}}$) in the vicinity of
the Fermi level, which in case of the raw data would require a complicated symmetrization procedure based on Fermi
surface fitting, being a source of additional errors. Finally, it provides the Green's function in absolute units,
allowing for quantitative comparison with other experiments and theory, even though the spectral function originally
measured by ARPES lacks the absolute intensity scale. Thereupon, we find the proposed analytical expressions to be
better estimates for the self-energy and both Green's functions and therefore helpful in calculations where comparison
to the experimentally measured spectral function is desirable.

\begin{figure}[b] \vspace{-0.5em}
\includegraphics[width=\columnwidth]{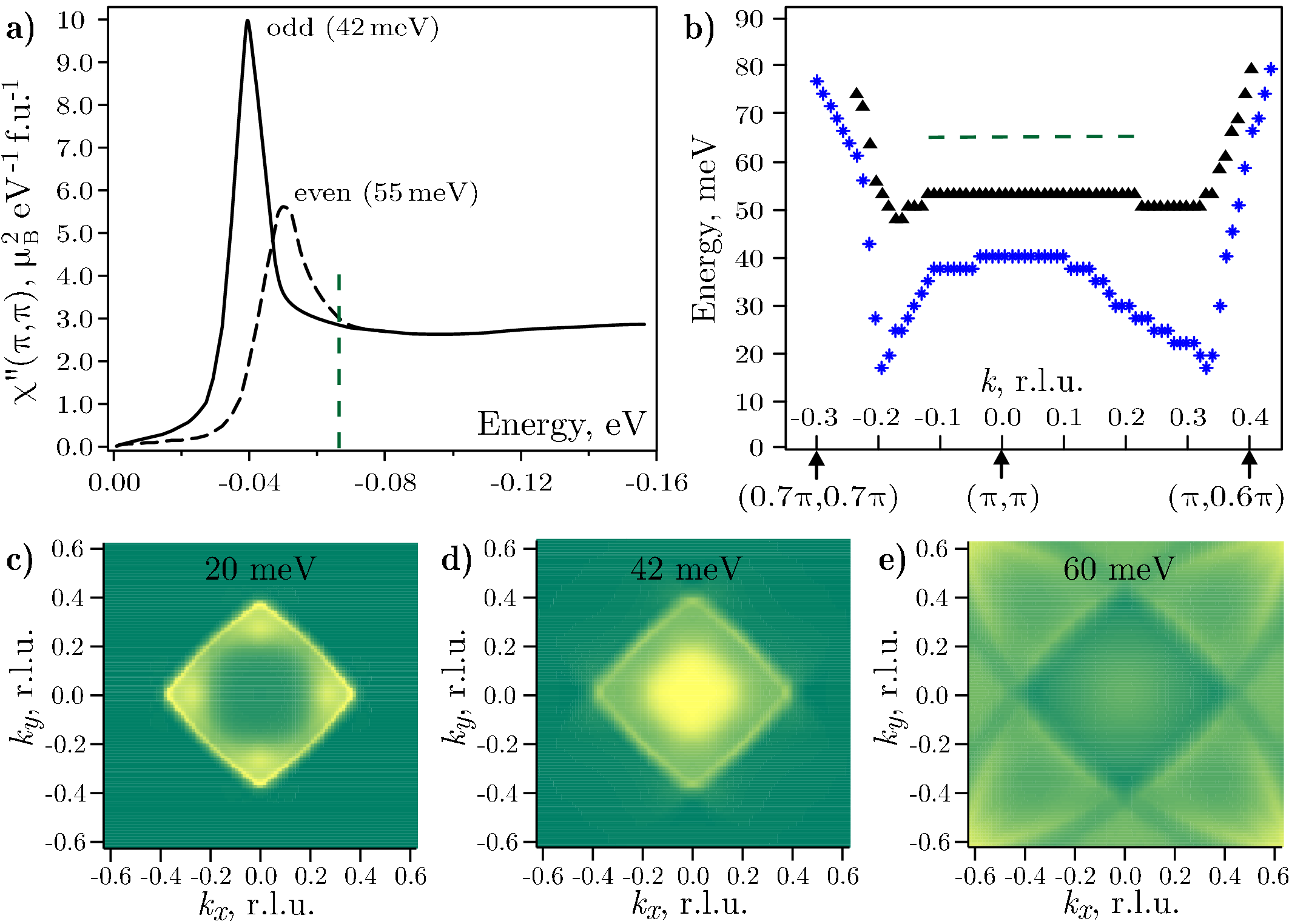}
\caption{\!\!\!\ColorOnline~\textbf{a)} $\textbf{k}$-integrated $\chi''\kern-0.3em=\kern-0.1em\operatorname{Im}(\chi)$
in the odd (solid curve) and even (dashed curve) channels. \textbf{b)}~$\textbf{k}$-dependence of the resonance energies
in odd (\lower0.27em\hbox{\textcolor{DarkBlue}{\textbf{*}}}) and even (\raise0.1em\hbox{$\scriptstyle\blacktriangle$})
channels along the high-symmetry directions $(0,0)$\,--\,$(\pi,\pi)$\,--\,$(\pi,0)$. The dashed lines mark the onset of
the particle-hole continuum (position of the ``step'' in $\chi''_0$). Second row:~Constant energy cuts of $\chi''$ in
the odd channel below the resonance (\textbf{c}), at the resonance energy (\textbf{d}), and above the resonance
(\textbf{e}). The center of each BZ image corresponds to the $(\pi,\pi)$ point.} \label{f:Chi}
\end{figure}

Now we will apply the described model to calculating the dynamic spin susceptibility. Starting from the model data set
built for optimally doped BSCCO at 30\,K, with the maximal SC gap of 35~meV, we have calculated the Lindhard function
(Eq.\,\ref{EqHi0}) in the energy range of $\pm$0.25~eV in the whole BZ for the odd and even channels of the spin
response (see Fig.\,\ref{f:Chi0}a). To demonstrate that the contribution of the anomalous Green's function is not
negligibly small, in Fig.\,\ref{f:Chi0}b we show separately the normal and anomalous components of
$\chi_0^\textup{odd}$.

After that we calculated $\chi$ (Eq.\,\ref{EqRPA}) by adjusting the $J_\|$ and $J_\perp$ parameters to obtain correct
resonance energies at $(\pi,\pi)$ in the odd and even channels (42 and 55~meV respectively), as seen by INS in
\mbox{BSCCO} \cite{FongBourges99, HeSidisBourges00, CapognaFauque06}. The resulting $\chi^\textup{o,\,e}$ are
qualitatively similar to those obtained for the bare Green's function \cite{EreminMorr06}. The intensity of the
resonance in the even channel is approximately two times lower than in the odd channel, which agrees with the
experimental data \cite{PailhesUlrich06,CapognaFauque06}. On the other hand, for $J_\perp\kern-0.3em=\kern-0.1em0$ the
splitting between odd and even resonances does not exceed 5\,--\,6~meV, which is two times less than the experimental
value. This means that the out-of-plane exchange interaction (in our case $J_\perp/J_\|\approx0.09$) is significant for
the splitting and the difference in $\chi_0$ alone between the two channels cannot fully account for the effect.

In Fig.\,\ref{f:Chi}a we show both resonances, momentum-integrated all over the BZ. Here we would like to draw the
reader's attention to the absolute intensities of the resonances. A good estimate for the integral intensity in this
case is the product of the peak amplitude and the full width at half maximum, which for the odd resonance results in
0.12~$\mu_B^2/$f.u. in our case. This is in good agreement with the corresponding intensity in latest experimental
spectra on YBCO ($\sim$\,0.11~$\mu_B^2/$f.u.) \cite{WooNature06}.

As for the momentum dependence of $\chi$, Fig.\,\ref{f:Chi}b shows the dispersions of incommensurate resonance peaks in
both channels along the high-symmetry directions, calculated from the Green's function model with the self-energy
derived from the ARPES data. We see the W-shaped dispersion similar to that seen by INS on YBCO
\cite{HaydenNature04,Tranquada05} and to the one calculated previously by RPA for the bare Green's function
\cite{EreminMorr05, EreminMorr06}. At $(\pi,\pi)$ both resonances are well below the onset of the particle-hole
continuum at $\sim\,$65\,meV (dashed line), which also agrees with previous observations \cite{PailhesSidis04,
EreminMorr05, EreminMorr06}. At higher energies magnetic excitations are overdamped, so the upper branch of the
``hourglass'' near the resonance at $(\pi,\pi)$ suggested by some INS measurements
\cite{HaydenNature04,PailhesSidis04,Tranquada05} is too week to be observed in the itinerant part of $\chi$ and is
either not present in BSCCO or should originate from the localized spins.

In Fig.\,\ref{f:Chi} we additionally show three constant-energy cuts of $\chi$ in the odd channel below the resonance,
at the resonance energy, and above the resonance. As one can see, besides the main resonance at $(\pi,\pi)$ the
calculated $\chi$ reproduces an additional incommensurate resonance structure, qualitatively similar to that observed in
INS experiments \cite{HaydenNature04}. Below the resonance the intensity is concentrated along the $(k,0)$ and $(0,k)$
directions, while above the resonance it prevails along the diagonal directions $(k,\pm k)$.

In this work we have demonstrated the basic relationship between the ARPES and INS data. The comparison supports the
idea that the magnetic response below $T_\textup{c}$ (or at least its major constituent) can be explained by the
itinerant magnetism. Namely, the itinerant component of $\chi$, at least near optimal doping, has enough intensity to
account for the experimentally observed magnetic resonance both in the acoustic and optic INS channels. The energy
difference between the acoustic and optic resonances seen in the experiments on both BSCCO and YBCO, cannot be explained
purely by the difference in $\chi_0$ between the two channels, but requires the out-of-plane exchange interaction to be
additionally considered. In this latter case the experimental intensity ratio of the two resonances agrees very well
with our RPA results. Also the calculated incommensurate resonance structure is similar to that observed in the INS
experiment. Such quantitative comparison becomes possible only if the many-body effects and bi-layer splitting are
accurately accounted for. A possible way to do that is to use the analytical expressions for the normal and anomalous
Green's functions proposed in this paper. We point out that such method is universal and can be applied also to other
systems with electronic structure describable within the self-energy approach.

We thank N.\,M.\,Plakida for helpful discussions. This~project~is~part~of~the~Forschergruppe~FOR538 and is supported by
the DFG under Grants No. KN393/4. The experimental data were acquired at the Berliner
Elektronenspeicherring-Gesellschaft f\"{u}r Synchrotron Strahlung m.b.H.

\end{document}